# Uniting Superhydrophobic, Superoleophobic and Lubricating Fluid Infused Slippery Behaviour on Copper Oxide Nano-structured Substrates

Sanjeev Kumar Ujjain[$], Pritam Kumar Roy[$], Sumana Kumar, Subhash Singha and Krishnacharya Khare[*]

Department of Physics, Indian Institute of Technology Kanpur, Kanpur - 208016, India

[*]kcharya@iitk.ac.in

Keywords: Superhydrophobic, Superoleophobic, Lubricant infused slippery, Nanostructures, Metal oxides

**Abstract**

Copper oxide nanostructures with spherical (0D), needle (1D) and hierarchical cauliflower (3D) morphologies are used to demonstrate superhydrophobic, superoleophobic and slippery behavior. These nanostructures are synthesized on galvanized steel substrates using a simple chemical bath deposition method by tuning precursor concentration. Subsequent coating of low surface energy polymer, polydimethylsiloxane, results in superhydrophobicity with water contact angle ~160(2)° and critical sliding angle ~2°. When functionalized with low-surface energy perfluoroalkyl silane, these surfaces display high repellency for low surface tension oils and hydrocarbons. Among them, the hierarchical cauliflower morphology exhibits better re-entrant structure thus show the best superoleophobicity with 149° contact angle for dodecane having surface tension 25.3 mNm$^{-1}$. If these nanostructured substrates are infused with lubricant Silicone oil, they show excellent slippery behavior for water drops. Due to the lubricating nature of Silicone oil, the Silicone oil infused slippery surfaces (SOIS) show low contact angle hysteresis (~2°) and critical tilt angle (~2°). The hierarchical cauliflower nanostrcuture exhibit better slippery characteristics and stability compared to the other nanostructured surfaces.

**Introduction**

Past decade has attracted tremendous scientific interest in superhydrophobic surfaces because of their potential applications in many fields, such as transport of microdroplets, biochemical separation, drug delivery, tissue engineering, anticorrosion, self-cleaning, drag-reduction coating and microfluidic lab-on-chip devices.[1-8] Inspired by the investigation on lotus leaf effect, synthetic superhydrophobic surfaces are fabricated on a variety of substrates including metals[9], glass[10], polymers[11] and fabrics[12] by combining hierarchical nano- and microstructures along with low surface energy fluorinated molecules, resulting in very high water contact angle (CA > 150°) and a very low sliding angle (SA < 10°). [13, 14]

In addition to repelling water, superoleophobic surfaces repel organic liquids (oils/hydrocarbons) with low surface tensions and thus create the surface resistant against organic contamination.[15] However, in contrast to superhydrophobicity, achieving superoleophobicity entails a second essential feature related to a very specific surface morphology; i.e., re-entrant/overhanging surface features.[16] Generally, fabrication of surfaces with diminished wettability relies on roughness of textured surface as all non-textured (smooth) surfaces, regardless of their chemical compositions, are intrinsically oleophilic.[16] Young's relation for determining wettability (contact angle $\theta_Y$) of a liquid on smooth surface is given by:

$$\cos\theta_Y = (\gamma_{SV} - \gamma_{SL})/\gamma_{LV} \qquad (1)$$

Here $\gamma$ denotes the interfacial surface tension, and $S$, $L$, and $V$ stands for solid, liquid and vapour phase, respectively. Non-textured surfaces, if modified with low surface energy fluorosilane molecules (one with lowest surface energy reported is $\gamma_{SV}$ = 6 mJ/m$^2$ [17]) result in contact angle ($\theta_Y$) < 90° for oils. For instance, hexadecane (surface tension $\gamma_{LV}$ = 27.6 mJ/m$^2$) showed $\theta_Y$ ~ 80° while for water (surface tension $\gamma_{LV}$ = 72.1 mJ/m$^2$) $\theta_Y$ ~120°.[18] In contrast, textured surfaces modified with fluorinated silanes, on several occasions have demonstrated $\theta$ > 150° for oils and $\theta$ > 160° for water[16] with very low roll-off angle ($\omega$) as the liquid droplets can be stabbed on the top of the roughness asperities due to pockets of air trapped underneath[19], as discussed by Cassie-Baxter relation:

$$Cos\theta^* = -1 + f(1 + \cos\theta_{smooth}) \qquad (2)$$

where $\theta^*$ is defined as apparent contact angle on textured (rough) surface and $f$ represents the fraction of area at the solid-liquid interface. The relation (2) implies that ultimately, if $f \rightarrow 0$ then $\theta^* \rightarrow 180°$.

Although, this approach while promising, it suffers from inherent limitations related irreversible defects arising during fabrication and mechanical damage which enhance pinning effect in the superhydrophobic surface and stop liquid mobility.[20] In order to overcome these limitations, taking inspiration from Nepenthes pitcher plants [21], stable, defect-free and inert 'slippery' interface have been developed by lubricating liquid-infused porous surface in which micro-, nanostructures locked the infused lubricant in place.[20] Lubricating fluid is intrinsically smooth and defect-free affords instantaneous self-repair by wicking into dent sites in the underlying substrate capable of repelling various liquids[20] and ice.[22] Recently, Anand et al. has demonstrated reduced pinning of the condensate droplets by a hierarchical micro-nano textured surface by impregnating with an appropriate lubricant[23] while, Li et al. suggested self cleaning properties of hydrophobic liquid-infused porous poly(butyl methacrylate-co-ethylene dimethacrylate) surface.[24]

Therefore, it would be worthwhile to notice that combination of the nano/micro, hierarchical and re-entrant structures along with well-matched solid and liquid surface energy is the most crucial parameters to create highly stable superhydrophobic, superoleophobic and slippery surfaces but making such robust textured roughness is challenging.[20] Although current state-of-the art have generated efficient surfaces with precise control of the nano and micro structures, however reported fabrication processes for the creation of surface roughness involving lithographic means[20, 25], micro-fabrication[26-28], self assembly[29, 30], or by the use of polyelectrolyte multilayers (PEMs) assembled by the layer-by-layer technique[31], sol-gel methods[32], spin-coating[33], electrochemical deposition which requires a conductive substrate[34], and polymer imprinting[2, 35-40] are sophisticated and costly, and cannot be implemented on a large scale. The complications are mostly associated with the manufacturing of a robust hierarchical structure with re-entrant and convex morphology which is the key for superoleophobicity.

Steel considered to be the workhorse of our society as the most essential engineering material in field of construction, food, petrochemical, maritime and aviation industries.[41] Its broad applications can be further augmented by making it super-repellent for water/oil/hydrocarbons, especially for industries where metal−fluid contact is common. The super-repellency provides resistance to antifouling/corrosion properties increasing their life-time and allows complete dewetting of transport channels (tanks/pipes), thereby sinking product loss due to residual surface wetting.[42]

In this work, we utilized chemical bath deposition (CBD) for synthesizing four different morphologies of copper oxide (CuO) nano-/micro and hierarchical structures on steel substrates. The CBD method involves very simple instrumentation facility and can be used to fabricate a single crystalline material such as metal oxides and hydroxides on a wide variety of substrates, because in thermodynamics equilibrium conditions each metal complex in the precursor solution is singly deposited on the substrate surface.[43] Furthermore, these CuO nano-/micro textured steel substrates were coated with polymethylsiloxane (PDMS) by dip coating resulting in highly robust superhydrophobic surfaces. The effect of micro-/nano structures on superhydrophobic, superoleophobic and lubricant based slippery behaviour have been studied. Moreover, the fabrication conditions involves in whole process are mild (low temperature, dilute solutions and air atmosphere) and fairly easy; required no sophisticated instrument.

**Experimental Section**

*Materials:* Poly(dimethylsiloxane) (PDMS) prepolymer (Sylgard 184A) and thermal curing agent (Sylgard 184B) were bought from Dow Corning Corp. 1H,1H,2H,2H-perfluorooctyltriethoxy silane ($C_{14}H_{19}F_{13}O_3Si$) (PTES) was purchased from Alfa Aesar. Silicone oil ($\eta$=350 cSt), copper sulphate ($CuSO_4$), L-ascorbic acid, sodium hydroxide (NaOH) and sodium borohydride ($NaBH_4$) were of analytical grade and purchased from Loba Chemie. Sodium dodecyl sulphate (SDS), ethanol and toluene were procured from Merck. Galvanized steel (GI) substrates (Tata Steel, India) (4cm×2cm×0.2cm) were ultrasonically cleaned in acetone and then rinsed with de-ionized water sequentially before use.

*Characterization:* The surface morphologies of the fabricated copper oxide particle on different substrates and the corresponding superhydrophobic PDMS films coated samples were investigated using field emission scanning electron microscopy (FESEM: JEOL, JXA-8230) at an accelerating voltage of 10 kV and RMS surface roughness was determines using atomic force microscopy (Park system XE-70) in a tapping mode in the range of scanning area of 3μm×3μm. Powder X-ray diffraction (XRD) measurements of copper oxide nanoparticles grown on steel substrates were performed with a X'Pert Pro MPD X-Ray Diffractometer. Static contact angle (CA) ($\theta_S$), Advancing contact angle ($\theta_A$) and Receding contact angle ($\theta_R$) were measured using Milli-Q water at room temperature with a Contact Angle Goniometer model: Data Physics (OCA 35). $\theta_S$ was measures using Laplase-Young fitting model of a 2 μl water droplet placed on the horizontal glass substrate. $\theta_A$ and $\theta_R$ were measured by adding and then withdrawing 2 μl of water drop respectively. The contact angle hysteresis ($\theta_A - \theta_R$) (CAH) was obtained using the circle fitting method. Static tilt angle ($\alpha$) was obtained by observing the minimum tilt angle required to move the water drop (10 µl) from horizontal surface. The velocity of the drop (10 µl) was measured by recording the movement of water drops on a tilted surface ($\alpha = 12^o$) and then calculating the distance travelled per unit time. All $\theta$ and $\alpha$ value are in the range of $\pm 2^o$. All the measurements were performed in duplicate on a given substrate and two such substrates for each composition. The values mentioned are average of these measurements.

*Synthesis of Copper Oxide nanoparticles:* Morphology controlled growth of copper oxide nanoparticles film on Fe substrate in aqueous solvent was typically processed by bath deposition method. Equal volume of two aqueous sodium dodecyl sulfate (SDS) solutions (10 mM), one containing cupric sulphate ($CuSO_4.5H_2O$) (10 mM–100 mM) and the other containing L-ascorbic acid (20 mM–1M) were mixed by magnetic stirrer and temperature was kept constant at 70$^o$C. Colour change occurred in the aqueous phase from blue to pale green and finally colourless. The pH of the solution was adjusted to be 11 using NaOH solution. On addition of NaOH solution, the colourless solutions change to yellow, reddish orange and reddish brown respectively, depending on the concentration of metal precursor solution. The cleaned steel substrates of dimension (4×2cm) were hanged in the above nucleated solution. After stirring for 10 min, 10 ml of (100 mM-1M) $NaBH_4$ aqueous solution was dropwise added into the reaction vessel. On addition, all the solutions gradually became reddish black. The reaction mixture was further stirred overnight in ambient atmosphere at 70$^o$C, to allow the reaction to complete. Thin film of CuO nanoparticles formed on steel were taken out and washed with ethanol and DI water to remove surfactant. The films were dried in air at 60$^o$C and used for further studies. For synthesizing CuO nanoparticles on PDMS film, glass and steel mess substrates, highest precursor concentration (i.e. 100 mM $CuSO_4$) was used.

*Preparation of PDMS coated superhydrophobic copper oxide textured steel:* Copper oxide nanoparticle textured steel surfaces are superhydrophilic. In order to make them superhydrophobic, PDMS was used as a low surface energy material for the nanostructures surface modification. Typically, PDMS prepolymer (4 gm) and thermal curing agent (2 gm) were dissolved in 200 ml toluene using ultrasonic bath for 30 min. Then copper oxide nanoparticle textured surfaces were dip coated a dip and lifting speed 200 mm/min and dried in air for 5 min. After drying, the samples were cured at 110°C for 30 min. Consequently, superhydrophobic PDMS coated steel surfaces were obtained.

*Surface Modification by Perfluoroalkyl Silane:* The PDMS coated copper oxide textured steel samples were exposed to a UV Ozone cleaner (Model No:ACS-40W6-UVO) for 10 min to oxidize the PDMS surface and generate oxygen functionalities.[44] These functionalities act as anchor sites for perfluoroalkyl silane grafting. In a typical process, the oxidized samples were placed in a vacuum chamber and 20 µl of 1H,1H,2H,2H-perfluorooctyltriethoxysilane ($C_{14}H_{19}F_{13}O_3Si$) (PTES) deposited on a glass slide was placed at a sufficient minimum distance from samples. Under low pressure, PTES vapour was deposited on samples for 20 min.

*Silicone oil infused slippery surface:* Slippery surfaces based on PDMS coated copper oxide textures steel surfaces infused with Silicone oil were prepared by immersing the steel samples in Silicone oil for 2 min and then picked out with a speed of 200 mm/min. The porous PDMS absorb Silicone oil and excess oil was removed by gravity drainage by hanging samples for 30 min. After drainage, the steel surfaces were spinned at 500 rpm for 30 sec to maintain uniformity in the oil film thickness.

**Results and Discussion**

**Structure, Morphology and Roughness**

Morphology controlled synthesis of nanocrystals with well defined shape and uniform size has been achieved by several methods involving conventional solid state process and wet synthetic routes, such as hydrolysis, pyrolysis, precipitation, and hydrothermal/solvothermal.

Among all these methods, the solution assisted synthesis by chemical bath deposition (CBD) of the precursor may be the most facile and effective approach to develop nanocrystals at relatively low temperatures, which is exempted from post calcination.

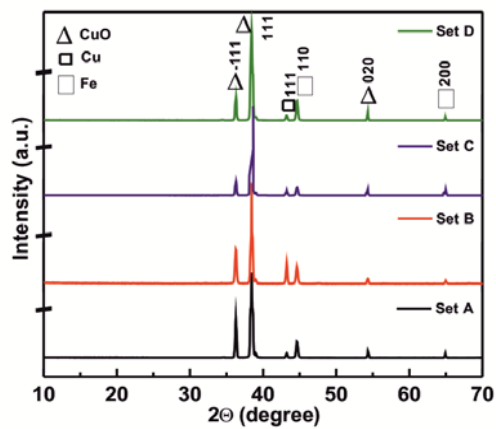

Figure 1. XRD patterns of CuO nanoparticles on steel substrates obtained by the chemical bath deposition method using different concentration (Set A) 10 mM, (Set B) 20 mM, (Set C) 50 mm and (Set D) 100 mM aqueous $CuSO_4$ solution in the presence of the reducing agents ascorbic acid and sodium borohydride respectively at pH 11.

Besides this, CBD exhibits considerable influence of metal salt precursor ($CuSO_4$) concentration on the final structure and morphology of the as-prepared CuO patterns on steel substrate. All obtained CuO samples are of Base-centered monoclinic structure, space group: C2/c(15) (JCPDS Card No. 001-1117, a = 4.653 Å, b = 3.41 Å, c = 5.108 Å, b = 99.48°). The XRD pattern shown in Figure 1 exhibits peak at $2\theta$: 35.74, 38.95, 53.88 corresponding to Miller indices (-111), (111) and (020), confirming CuO. Very diminished peak at 43.3° corresponds to (111) plane of Cu (JCPDS Card No. 01-070-3039). Presence of two characteristic peaks for Fe at $2\theta$: 44.65 and 64.98 corresponding to (110) and (200) Miller indices confirmed the presence of Fe substrate underneath CuO.

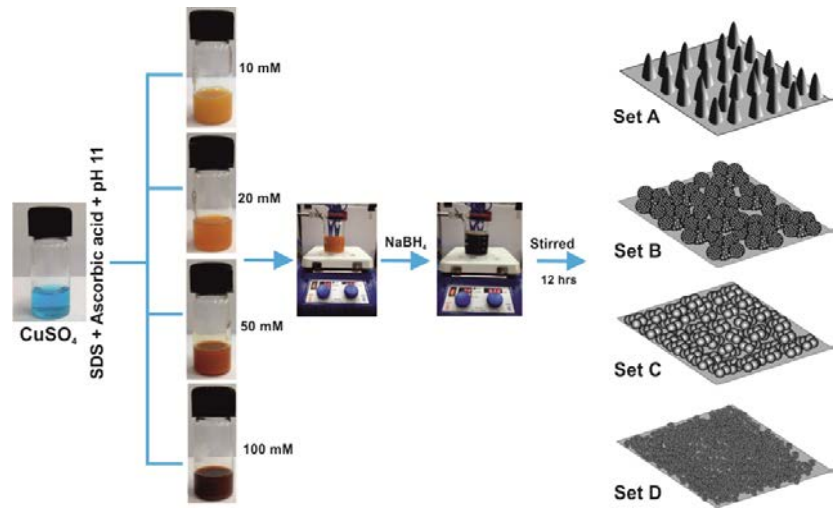

Figure 2. Schematic illustrating precursor concentration dependent CuO nano-/micro particles growth on steel substrates by chemical bath deposition method.

The summarized reaction conditions and morphologies are illustrated in Figure 2 and the representative SEM images are shown in Figure 3. The precursor solution for the crystallization of CuO was prepared using deionised water, $CuSO_4$, surfactant SDS, L-ascorbic acid, NaOH and $NaBH_4$ solution in the order depicted above. Depending on the concentration of aqueous SDS $CuSO_4$ solution (blue) from 10 to 100 mM, they demonstrate change in colour from yellow to wine reddish on addition of L-ascorbic acid at constant pH 11. After adding reducing agent $NaBH_4$, the crystallization of CuO of different morphology took place. The increase in concentration of $CuSO_4$ induced a reduction in crystallite size probably due to enhancement of the nucleation rate which enhanced growth kinetics of nanocrystals. Field emission scanning electron microscope (FE-SEM) images clearly demonstrate the delicate morphology control that can be achieved by adjusting the concentration of precursor solution. 10mM $CuSO_4$ solution results in needle-like morphology with height around 1 µm, (Figure 3a & e) which are almost perpendicular to the substrate. The images show that the needles are very sharp, with tip diameter in the range of tens of nanometer. Upon increasing the $CuSO_4$ concentration to 20 mM, hierarchical cauliflower like morphology is obtained with an average diameter of about 0.5 to 2 µm (Figure 3b). Higher magnification image (Figure 3f) reveals that these hierarchical cauliflowers are also composed of nano-flakes.

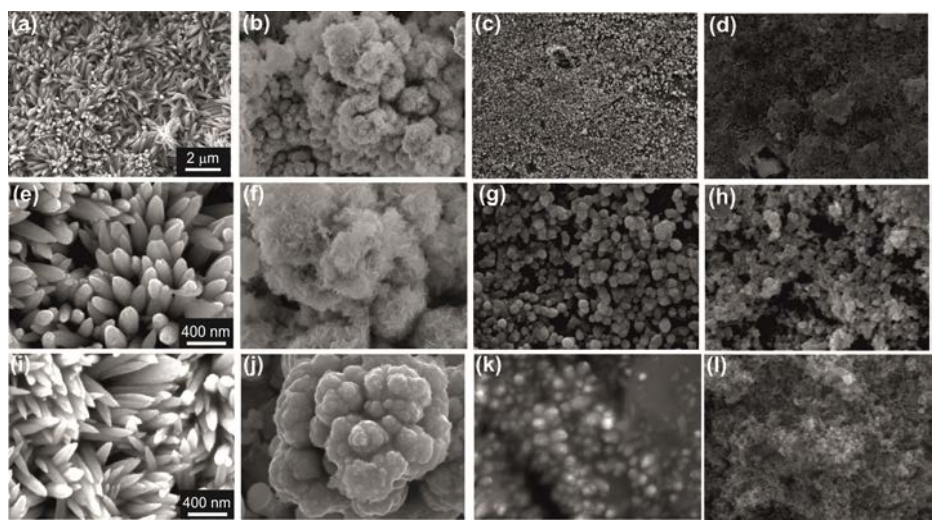

Figure 3. FESEM micrographs of CuO obtained by the chemical bath deposition process varying concentration of CuSO$_4$ precursor solution. (a)&(e) Nano-needle(10 mM), (b)&(f) Hierarchical cauliflower (20 mM), (c)&(g) nanosphere (50 mM) and (d)&(h) nanosphere cluster (100 mM). Micrograph (i) to (l) showing PDMS coated surfaces revealing no significant change in morphology.

Figure 3c & g display spherical morphologies with an average diameter 80 nm obtained from 50 mM precursor solution. If concentration is further increased to 100 mM, CuO nanospheres get agglomerated resulting in larger clusters (Figure 3d & h).

In order to reduce surface energy and improve mechanical stability, the hydrophilic CuO nano-patterns were dip coated in dilute PDMS solution followed by curing at 110°C for 30 minutes. The resulting CuO patterned surfaces shown in Figure 3 i-l, indicating no change in overall morphology. Here we should note nanoscale roughness in hierarchical cauliflower and nanosphere morphologies is smeared out due to PDMS filling, but we will later show that this does not affect any of the physical behavior of the system.

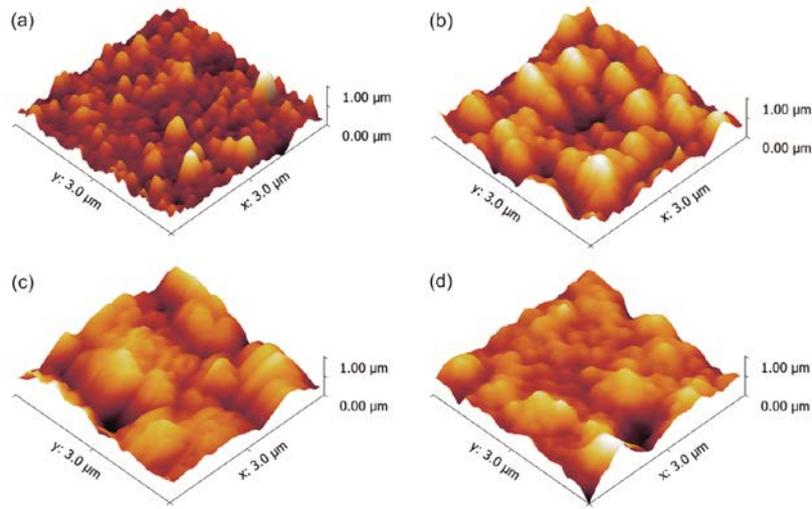

Figure 4. Atomic force microscopy (AFM) images of PDMS coated CuO film on steel substrates. The root mean square (RMS) roughness values of CuO films were (a) 127.7 nm (nano-needle), (b) 164.5 nm (hierarchical cauliflower), (c) 143.7 nm (nanosphere) and (d) 132.9 nm (nanosphere cluster).

Mechanical stability of the nano patterns, prior to and post PDMS coating, was studied by a peel test. Scotch tape (3M Magic™ Tape) with pressure-sensitive adhesive was applied on the surfaces, pressed thoroughly and peeled-off. The removed nanoparticles from the CuO nano-needle textured surface without PDMS coating and with PDMS coating were examined by SEM. Dense coverage of nano structures were found on the pealed tape from the surfaces without PDMS coating, while the coated surfaces didn't reveal much, indicating enhanced stability of nano patterns post PDMS coating. Furthermore, the surface roughness of the PDMS coated CuO textured surfaces were determined by AFM. Figure 4a-d shows corresponding 3D AFM images after PDMS coating providing rms roughness of 127.7 nm, 164.5 nm ,143.7 nm and 132.9 nm of the nano patterns respectively. Subsequently, these PDMS coated CuO nano-textured steel surfaces with different morphologies and roughness were investigated for their wetting behaviour under different conditions.

**Superhydrophobicity Self-Cleaning Behavior**

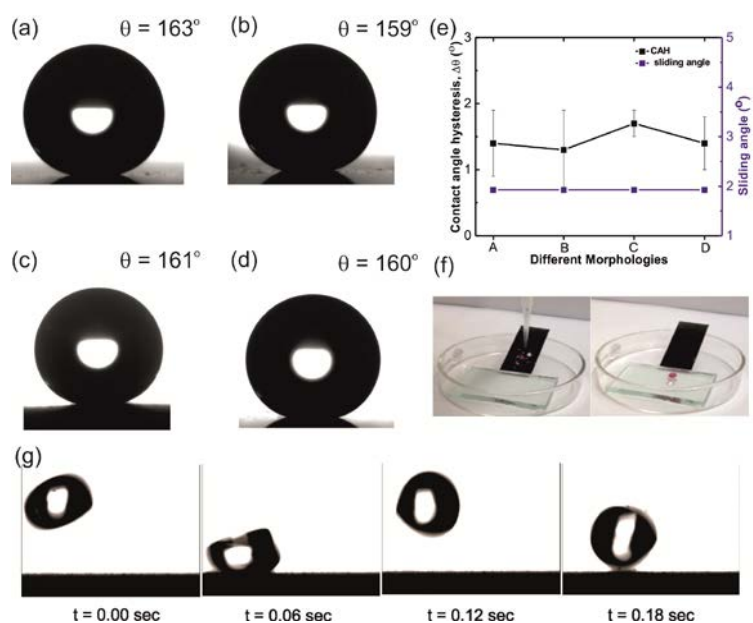

Figure 5. Contact angle on samples with various morphologies: (a) nano-needle, (b) hierarchical cauliflower, (c) nanosphere and (d) nanosphere cluster, (e) Contact angle hysteresis and sliding angle plots for different morphologies, (f) self-cleaning action of PDMS coated CuO textured steel substrate against various contaminants and (g) water drop bouncing off the superhydrophobic surface.

CuO nano-textured PDMS coated steel surface possess large roughness due to presence of CuO of different morphologies and low surface energy due to PDMS, which is essential to achieve superhydrophobicity.[16] All four CuO nano-textured morphologies after PDMS coating exhibit prominent superhydrophobicity with water contact angles as high as 163° (Figure 5a-d). The extreme water repellency of each surfaces also reflect very low contact angle hysteresis ($\Delta\theta \sim 2°$) and very low sliding angles ($\alpha \sim 2°$) for 2μl droplet volume (Figure 5e). The ultra-low contact angle hysteresis (difference between the advancing and receding contact angle of the droplet) and sliding angle (minimum surface tilt on which droplet starts moving) of these surfaces confirm uniformity and lack of pinning sites[20] which allowed water droplets to easily roll-off and bounce on them (Figure 5g; supplementary MovieS1).

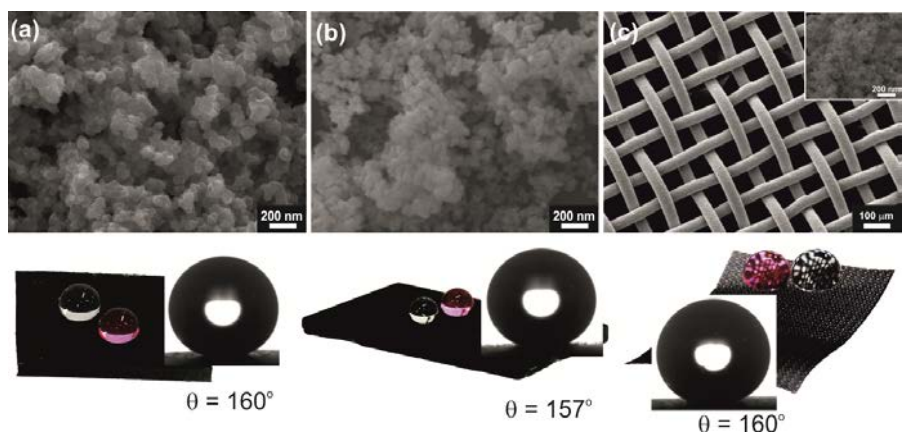

Figure 6. FESEM micrograph of CuO nanoparticles formed on (a) PDMS film, (b) glass and (c) steel mess (Inset showing high magnification image). Their respective digital photograph with water droplets sitting on the surfaces and contact angle is shown below them.

This superhydrophobic nature also protect the surfaces from wide range of contaminants by self-cleaning action which allows water droplets to collect and remove the contaminants from surfaces upon roll off (Figure 5f; supplementary MovieS2). To demonstrate versatility of CBD based superhydrophobic coatings on variety of substrates, CuO nanopatterns were grown on glass, PDMS sheet and steel mesh follwed by PDMS coating. All these substrates showed excellent superhydrophobicity (water contact angle ~ 160°) and mechanical stability (Figure 6).

**Superoleophobicity**

To generate repellency towards low surface tension oils and hydrocarbons, the PDMS coated nano-textured surfaces were further functionalized with a low-surface energy perfluoroalkyl silane. Each of these surfaces demonstrate extreme liquid repellency with contact angles 160° to 125° against liquids of surface tension ranging from 64.0 mNm$^{-1}$ (glycerol) to 25.3 mNm$^{-1}$ (dodecane) and depending on different morphologies (Figure 7a).

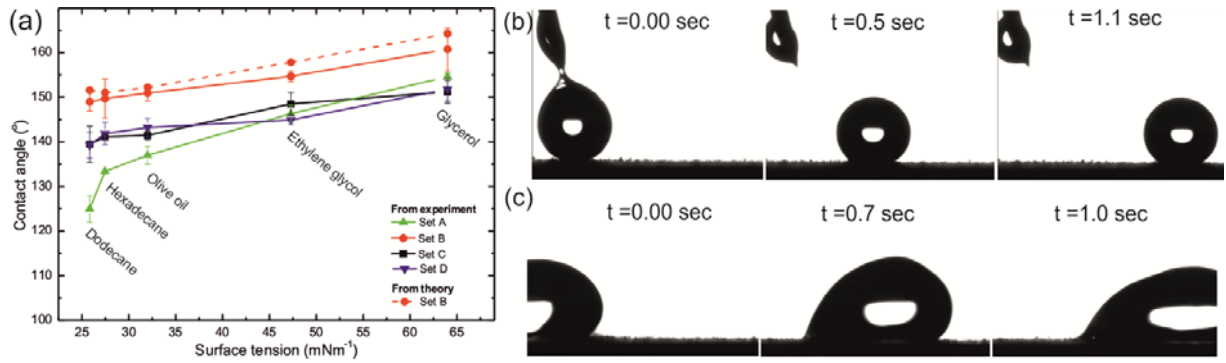

Figure 7. (a) Comparison of contact angle as a function of surface tension of test liquids on various morphologies of perfluoroalkyl silane functionalized PDMS coated CuO on steel substrates (Set A) nano-needle, (Set B) hierarchical cauliflower, (Set C) nanosphere and (Set D) nanosphere cluster. Their corresponding theoretical values are shown with dotted lines. Snaps of rolling glycerol (b) and dodacane (c) drops on hierarchical cauliflower textured surface.

The surface with cauliflower morphology having hierarchical nano-/micro structure with highest roughness displays higher contact angles, 160° for glycerol, which decreased to 149° for dodecane (Figure 7 b&c; supplementary MovieS3 & MovieS4), while the nano-needle surface showed inferior repellency for low surface tension liquid, around 125° for dodecane. This is in agreement with previous reports discussing the ideal design parameters for superoleophobic surfaces having re-entrant geometry.[16] Hierarchically structured cauliflower surface, possessing re-entrant structure, can trap higher fraction of air at both the coarser and finer length scale showing extreme superoleophobicity while nano-needle structure surface has one scale of texture. Consequently, the former can support low surface tension liquids in the Cassie state resulting in superoleophobic surface. The substrates possessing a predominantly spherical textures (Set C and Set D) demonstrated intermediate superoleophobic behaviour between the hierarchical and one scale textured.

As the cauliflower shape possessing re-entrant structure and demonstrating the highest oleophobicity among the studied textures, we approximated the contact angles for different test liquids using equation (3). [15, 16]

$$\cos\theta^*_{sphere} = -1 + \frac{1}{D^*_{sphere}}\left[\frac{\pi}{2\sqrt{3}}(1+\cos\theta)\right]^2 \qquad (3)$$

where, $\theta^*$ is the apparent contact angle on the textured surface and $\theta$ is the equilibrium contact angle on a smooth surface of the same substrate, given by Young's equation (1). $D^*$ is the spacing ratio defining surface porosity and given by $D^*_{sphere} = \left[\frac{(R+D)}{R}\right]^2$ for spherical textured surface. Here, R defines the radius of the sphere and 2D is the inter-sphere distance. The theoretical values calculated using the above equation precisely matches the experimental data trend with highest contact angle $164°$ for glycerol and $151°$ for dodecane (Figure 6a). The higher theoretical values for hierarchically textured surface are in agreement with the previous works on designing superoleophobic surfaces.[16]

**Silicone oil infused slippery surfaces (SOIS)**

SOIS were designed taking three criteria in consideration: (1) affinity of lubricating liquid with solid surface, i.e. the lubricating liquid must spread completely on the substrate, (2) slipping test liquid must be immiscible with the lubricant and (3) requirement of hydrophobic solid substrate. The first requirement is satisfied by using silicone oil, as it spreads completely on PDMS films. Since silicone oil is known to swell crosslinked PDMS, it is advantageous for our system as the swelling increases its adhesion with PDMS coated substrates. Second criteria was satisfied by taking water as test liquid which shows immiscibility with silicone oil.

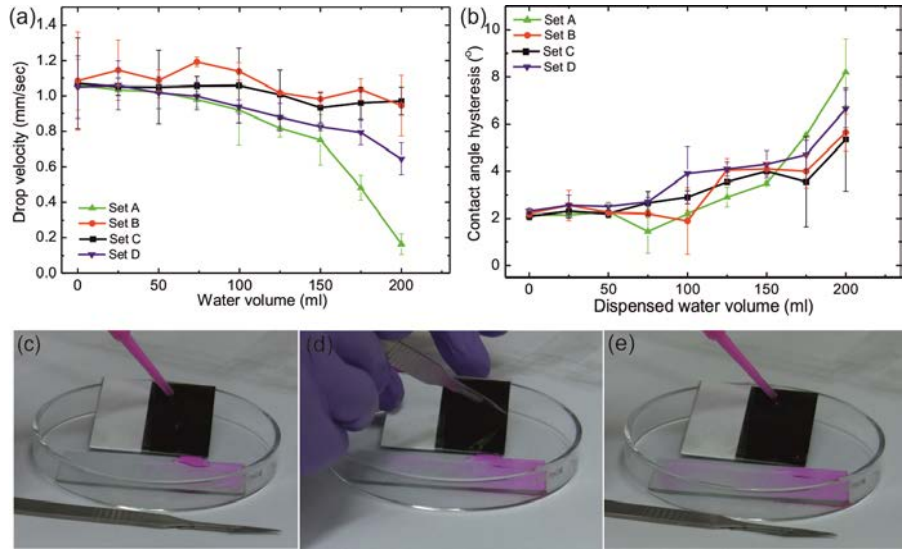

Figure 8. (a) Showing drop (10 µl) velocity with volume of water dispensed, (b) Contact angle hysteresis as a function of volume of water dispensed over four different morphologies of Silicone oil infused PDMS coated CuO textured steel surfaces. Snaps from movie demonstrating the fast recovery of the liquid-repellent property of a SOIS after critical physical damage. (c) showing coloured water drops sliding on hierarchical cauliflower morphology SOIS, (d) physically damaging film with knife and (e) water drops sliding after physical damage.

Third criteria is fulfilled automatically as PDMS coated substrates are inherently hydrophobic. Satisfying these criteria, we fabricated silicone oil infused nano textured slippery surfaces on steel substrates to slip water droplets upon tilting. Each of these SOIS demonstrates very low contact angle hysteresis ($\Delta\theta < 2°$) and low critical sliding angle ($\alpha < 2°$) and good slippery behavior with water drop velocity ($\geq 1.05$ mm/sec) for 10 µl droplet volume inclined at 12° (Figure 8).

Slippery behavior of the SOIS is found independent of underlying texture geometry provided sufficient lubricating fluid covers the entire surface. Static water drops deposited on silicone oil coated solid surfaces are cloaked with a thin layer of oil due to positive spreading coefficient of oil on water. Spreading parameter ($S_{OW(V)} = \gamma_{WV} - \gamma_{WO} - \gamma_{OV}$, subscript O, W and V represent oil, water and vapour respectively) for our system is 8.5 mN/m.[45] Once these oil cloaked water drops slip from the lubricant coated surface, they slowly remove oil from the surface thus decrease the lubricating layer thickness which affect the velocity of slipping water drop. We studies the effect of CuO nano morphology on the degradation of slippery behavior as a function of slipping water volume (Figure 8 a). It is clear from the figure that set B (hierarchical cauliflower) shows smallest degradation in slip velocity upon slipping water drops. This is due to the fact that the hierarchical cauliflower structures posses re-entrant curvature and lubricating oil trapped in the re-entrant structures are very hard to be removed. Therefore these structures show least degradation upon water flow compared to other structures. On the other hand, degradation on Set A sample is found to be the most due to their needle like structure. This degradation in slippery behavior of SOIS is related to the decreased lubricant thickness, which is also reflected in their increasing contact angle hysteresis (Figure 8 b). Expectedly, set B shows lowest increase in the contact angle hysteresis where as set A shows the largest due to their underlying morphology. Self-healing ability of the SOIS is also checked by large area physical damage created on silicone oil film using a knife. Silicone oil lubricating film quickly heals the damage and restores the slippery behavior within fraction of seconds by filling the damaged void area by surface capillary action (Figure 8c-e; supplementary MovieS5).

**Conclusion**

In this article, we have demonstrated a novel and convenient method to synthesize CuO nano-/micro structures with spherical (0D), needle (1D) and hierarchical (3D) cauliflower morphologies on steel substrate using chemical bath deposition (CBD). The change in morphology from 0D to 3D is precursor concentration dependent. Another advantage of this method is that, it is substrate independent and can be formed on polymer film, glass and mess.

On polydimethylsiloxane (PDMS) coating, these nano-/micro textured surfaces formed robust self-cleaning superhydrophobic surfaces with water drops bouncing on them as a consequence of low contact angle hysteresis ($\Delta\theta \leq 2°$) and sliding angle ($\alpha < 2°$). Perfluoroalkyl silane grafting resulted in low surface energy nano / micro textured substrates showing repellency against various liquid with surface tension $\geq 25.3$ mNm$^{-1}$ (dodecane). The cauliflower morphology textured steel surface outperformed other morphologies in terms of different liquid repellency because its hierarchical surface. In order to eliminate the constrains related to the self-healing on physical damage, silicone oil infused slippery surfaces (SOIS) were formed on these nano / micro textured steel substrates. These SOIS demonstrated excellent slippery behavior for water with quick self-healing against physical damage. These results indicate that a suitable nano-/micro textured robust structures for desired application (superhydrophobic/superoleophobic/slippery) can be fabricated very easily on various substrates. We expect this low-cost, fast and convenient method will pave a new way for designing and fabricating robust textured surfaces for numerous applications in research and industrial field.

**Acknowledgements**

S. K. Ujjain and P. K. Roy contributed equally to this work. This research work was supported by Hindustan Unilever Limited, India and DST, New Delhi through its Unit of Excellence on Soft Nanofabrication at IIT Kanpur.